\documentclass{PoS}

\usepackage[utf8]{inputenc}
\usepackage[english]{babel}

\usepackage{amsmath}
 
\usepackage[comma,square,numbers,sort&compress]{natbib}
\usepackage{lineno}
\usepackage{float}
\usepackage{color}
\usepackage{booktabs}
\usepackage{multirow}
\usepackage{soul}
\usepackage{xspace}

\usepackage{array}
\usepackage{makecell}

\newcommand{\pt}{\ensuremath{p_{\rm T}}\xspace}
\newcommand{\mpt}{\ensuremath{\langle p_{\rm T} \rangle}\xspace}
\newcommand{\nch}{\ensuremath{N_{\rm ch}}\xspace}

\title{Particle production and flow-like effects in small systems}

\ShortTitle{Small systems}

\author{\speaker{Antonio Ortiz, for the ALICE, ATLAS, CMS and LHCb Collaborations}\\
        Instituto de Ciencias Nucleares, UNAM, Circuito Exterior s/n, CU, Deleg. Coyoac\'an, Ciudad de M\'exico 04510\\
        E-mail: \email{antonio.ortiz@nucleares.unam.mx}}

\abstract{
Particle production in small systems (pp and p--Pb collisions) has unveiled unexpected collective-like behavior. In this work an overview of the current investigation on the similarities between small systems and heavy-ion collisions is presented. Recent results from the experiments at the LHC are discussed. They include measurements of multi-particle correlations, as well as, identified particle production as a function of charged-particle multiplicity density, and more recently, as a function of transverse spherocity. 

}

\FullConference{7th Annual Conference on Large Hadron Collider Physics - LHCP2019\\
		20-25 May, 2019\\
		Puebla, Mexico}

\begin{document}

\section{Introduction}

The goal of the heavy-ion program is to understand the behavior of Quantum Chromo-Dynamics (QCD) at high temperatures and densities. Results at the Large Hadron Collider (LHC) confirmed the formation of a new form of matter characterized by deconfinement, which is compatible with the theoretically predicted Quark-Gluon Plasma (QGP)~\cite{Busza:2018rrf}. It has been found that the low transverse momentum ($\pt<2$\,GeV/$c$) hadron production and multi-particle correlation measurements are well-described by fluid dynamic models, yielding that matter properties are consistent with a perfect fluid that exhibits small shear viscosity over entropy ratio, $\eta/s$~\cite{Romatschke:2017ejr}. Other observables such as jet suppression are also well-described by models which assume the formation of a QGP~\cite{Qin:2015srf}. The main conclusions arose from comparisons of heavy-ion data with reference data, such as minimum-bias pp and p--A collisions, where no signatures of  jet quenching were observed. 

\begin{figure}[htb]
\centerline{
\includegraphics[width=0.5\columnwidth]{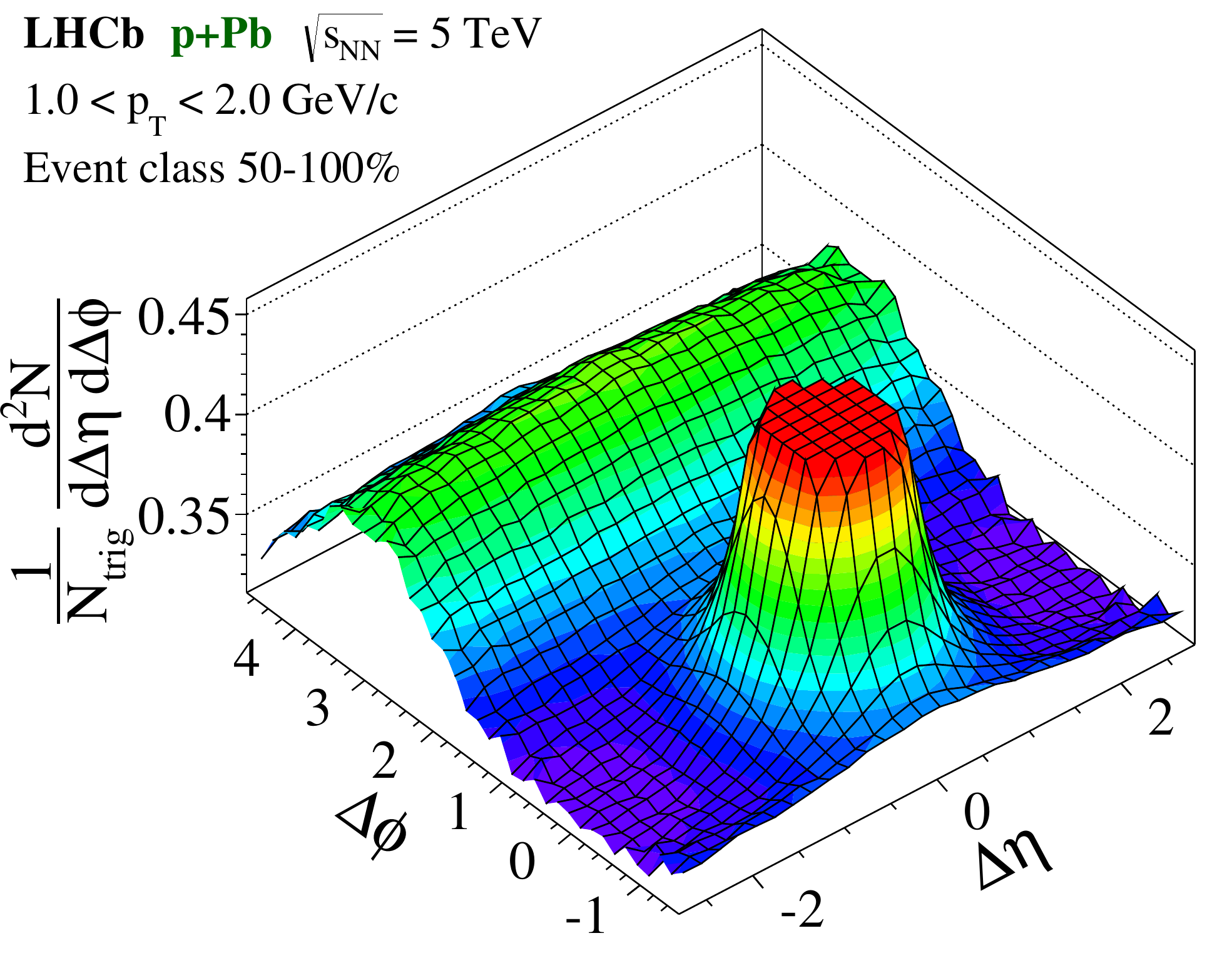}
\includegraphics[width=0.5\columnwidth]{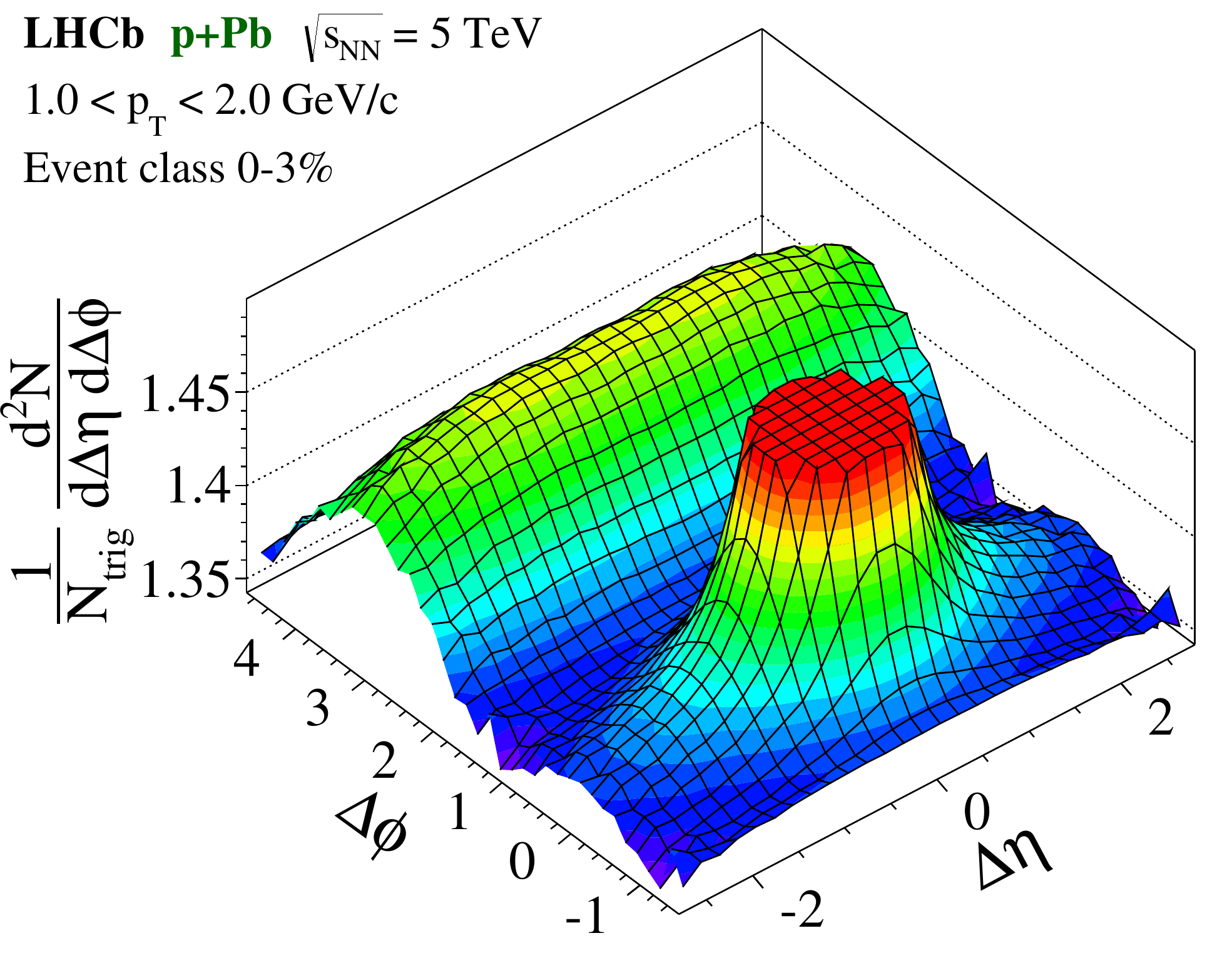}}
\caption{Two-particle correlation for (left) low- and (right) high-event activity in p--Pb collisions at $\sqrt{s_{\rm NN}}=5.02$\,TeV measured with the LHCb detector~\cite{Aaij:2015qcq}. Primary charged particles were selected in a \pt range of 1--2\,GeV/$c$. The near-side peak around $\Delta\eta=\Delta\phi=0$ is truncated in the histograms.}
\label{fig:1}
\end{figure}

The large amount of LHC pp data, allowed two-particle azimuthal correlations to be measured as a function of the charged-particle multiplicity (\nch). For the first time, very  similar  azimuthal anisotropies as in heavy-ion  collisions were observed in high-multiplicity pp collisions at $\sqrt{s}=7$\,TeV~\cite{Khachatryan:2010gv}. The analysis was further extended to lower and higher energies~\cite{Khachatryan:2016txc}, as well as other systems such as p--Pb at $\sqrt{s_{\rm NN}}=5.02$\,TeV~\cite{Abelev:2012ola,Aaij:2015qcq}. Figure~\ref{fig:1} shows the correlation for particles with $1 < \pt < 2$\,GeV/$c$ in low and very-high multiplicity p--Pb events. The long-range structure (often referred to as the near-side ridge) centred at $\Delta\phi=0$ and  elongated over the full measured $\Delta \eta$ range of 2.9 units is not present in the corresponding low-activity sample. Moreover, also for the first time, two long-range ridge-like structures, one on the near side and one on the away side, were observed when the per-trigger yield obtained in low-multiplicity events was subtracted from the one in high-multiplicity events. This suggested that the near-side ridge is accompanied by an essentially identical away-side ridge.

Naturally,  it  was  suggested  that  these anisotropies in small systems could have the same origin as in heavy-ion collisions, namely the hydrodynamic response  of  the  produced  medium  to  the  initial  shape  of the interaction region in the transverse plane~\cite{Bozek:2011if}. However, the main concern was on the applicability of hydrodynamics to small non-equilibrium systems. A big progress in this direction has been accomplished, for example using numerical holography, it has been demonstrated that hydrodynamics can describe the evolution of systems as small as $R\sim1/T_{\rm eff}$, $T_{\rm eff}$ being the effective temperature~\cite{Chesler:2015bba}. From the initial state perspective, the azimuthal anisotropy is due to initial state correlations present in the nuclear wave functions of the incoming nuclei~\cite{Schlichting:2016sqo}. The main question is whether azimuthal anisotropies established during the initial stages of the collision can survive subsequent final state interactions~\cite{Strickland:2018exs}. The third approach relies on partonic and hadronic transport models, for example AMPT~\cite{Lin:2004en}. This model qualitatively (and sometimes quantitatively) describes small system flow signals for various collision systems and energies. The big issue is that in contrast to  fluid dynamic simulation, its applicability relies on a sufficiently large mean free path, which is hard to reconcile with the idea of the strongly coupled hydrodynamic system.  

From the experimental side, one challenge for small collision systems is the strong correlation between multiplicity (sensitive to low-\pt particles) and hard physics (high-\pt particles)~\cite{Acharya:2019mzb}. It has been shown to be reduced if event multiplicity is determined in a pseudorapidity region far from that where the observable of interest is measured. However, an additional treatment of the unwanted particle correlations (from jets or resonances) has to be implemented in data analysis.  

It is clear that we are still far from fully understanding the similarities between small systems and heavy-ion data. A more detailed review on small systems can be found in Ref.~\cite{Nagle:2018nvi}.  In this proceedings, a discussion of recent results on identified particle production and  multi-particle correlations in small systems is presented in Sections 2 and 3, respectively. Finally, Section 4 presents a summary of the main results. 

\section{Identified particle production as a function of multiplicity}

For pp collisions at $\sqrt{s}=7$\,TeV, the \pt spectra of different particle species ($\pi^{\pm}$, ${\rm K}^{\pm}$, ${\rm K_{S}^{0}}$, ${\rm K^{*}(892)^0}$, p, ${\rm \overline{p}}$, $\phi(1020)$, $\Lambda$, $\overline{\Lambda}$, $\Xi^-$, ${\overline \Xi}^+$, $\Omega^-$ and ${\overline \Omega}^{+}$) have been measured as a function of the event multiplicity and over a broad \pt range~\cite{Acharya:2018orn}. In particular, at low \pt  ($\pt<3$\,GeV/$c$), all spectra are observed to become harder with increasing charged-particle multiplicity density, the effect is more pronounced for heavier particles. Moreover, the modification of \pt spectra with respect to the inclusive measurement follows a different pattern for mesons and baryons, except for resonances, which follow baryons at a low-\pt of up to approximately 2\,GeV/$c$ and tend to be modified similarly to mesons above a \pt of 2\,GeV/$c$. Furthermore, it was demonstrated that the evolution of the baryon/meson ratios as a function of $\langle {\rm d}N_{\rm ch}/{\rm d}\eta \rangle$ exhibits an universal pattern for all collision systems~\cite{Adam:2016dau}. This behavior might indicate a common mechanism at work that depends solely on final-state multiplicity density.

\begin{figure}[htb]
\centerline{
\includegraphics[width=0.5\columnwidth]{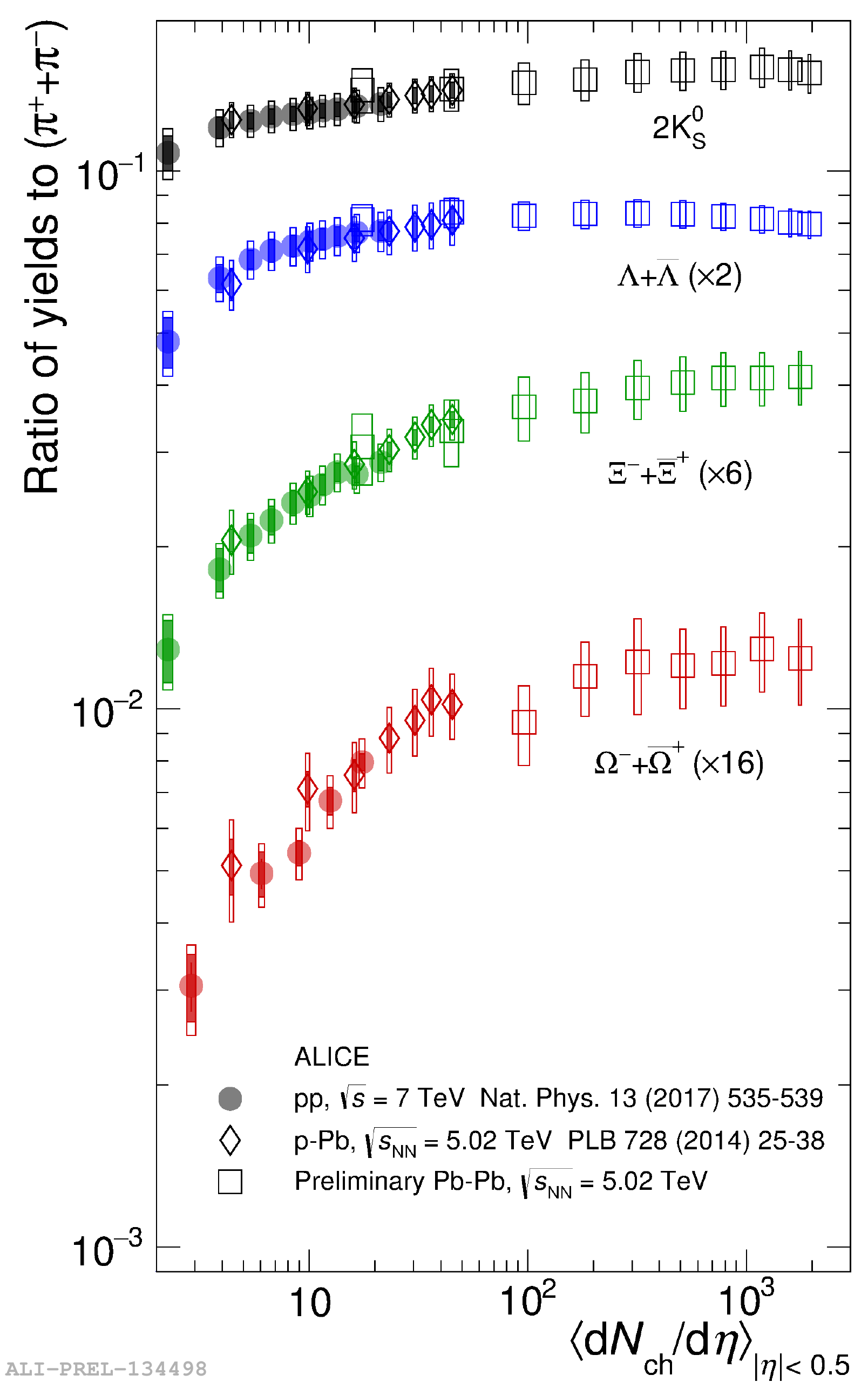}}
\caption{\pt-integrated yield ratios to pions ($\pi^{+}+\pi^{-}$) as a function of $\langle {\rm d}N_{\rm ch}/{\rm d}\eta \rangle$ measured in $|y|<0.5$. The error bars show the statistical uncertainty, whereas the empty and dark-shaded boxes show the total systematic uncertainty and the contribution uncorrelated across multiplicity bins, respectively. The values measured for pp collisions at $\sqrt{s}=7$\,TeV~\cite{Acharya:2018orn} are compared with results obtained in p--Pb~\cite{Abelev:2013haa} and Pb--Pb collisions at the LHC.}
\label{fig:4}
\end{figure}

Figure~\ref{fig:4} shows the ratios of the yields of ${\rm K^{0}_{S}}$,  $\Lambda$, $\Xi$ and $\Omega$ to the pion ($\pi^{+}+\pi^{-}$) yield as a function of $\langle {\rm d}N_{\rm ch}/{\rm d}\eta \rangle$ for pp, p--Pb and Pb--Pb collisions at the LHC energies. It is worth noting that the ALICE results for small systems~\cite{ALICE:2017jyt} are now compared with the latest data from Pb--Pb collisions at $\sqrt{s_{\rm NN}}=5.02$\,TeV.  A significant enhancement of strange to non-strange hadron production is observed with increasing particle multiplicity in pp collisions. The behavior observed in pp collisions resembles that of p--Pb collisions at a slightly lower centre-of-mass energy, in terms of both the values of the ratios and their evolution with multiplicity.  Strangeness production in heavy-ion collisions can be described in the framework of statistical-hadronisation models~\cite{Cleymans:2006xj,Andronic:2008gu}. While in central heavy-ion collisions, the yields of strange hadrons is consistent with the expectation from a grand-canonical ensemble; the strange hadron yields in elementary collisions, such as pp and ${\rm e^{+}+e^{-}}$, are suppressed with respect to the predictions of the (grand-canonical) thermal models. In the strangeness-canonical approach, within the framework of the THERMUS~\cite{Wheaton:2004qb} statistical hadronisation model, the  multiplicity and system size dependence of all measured light  flavour hadrons can be described,  within 2 standard deviations~\cite{Vislavicius:2016rwi}.  

\begin{figure}[htb]
\centerline{
\includegraphics[width=1.0\columnwidth]{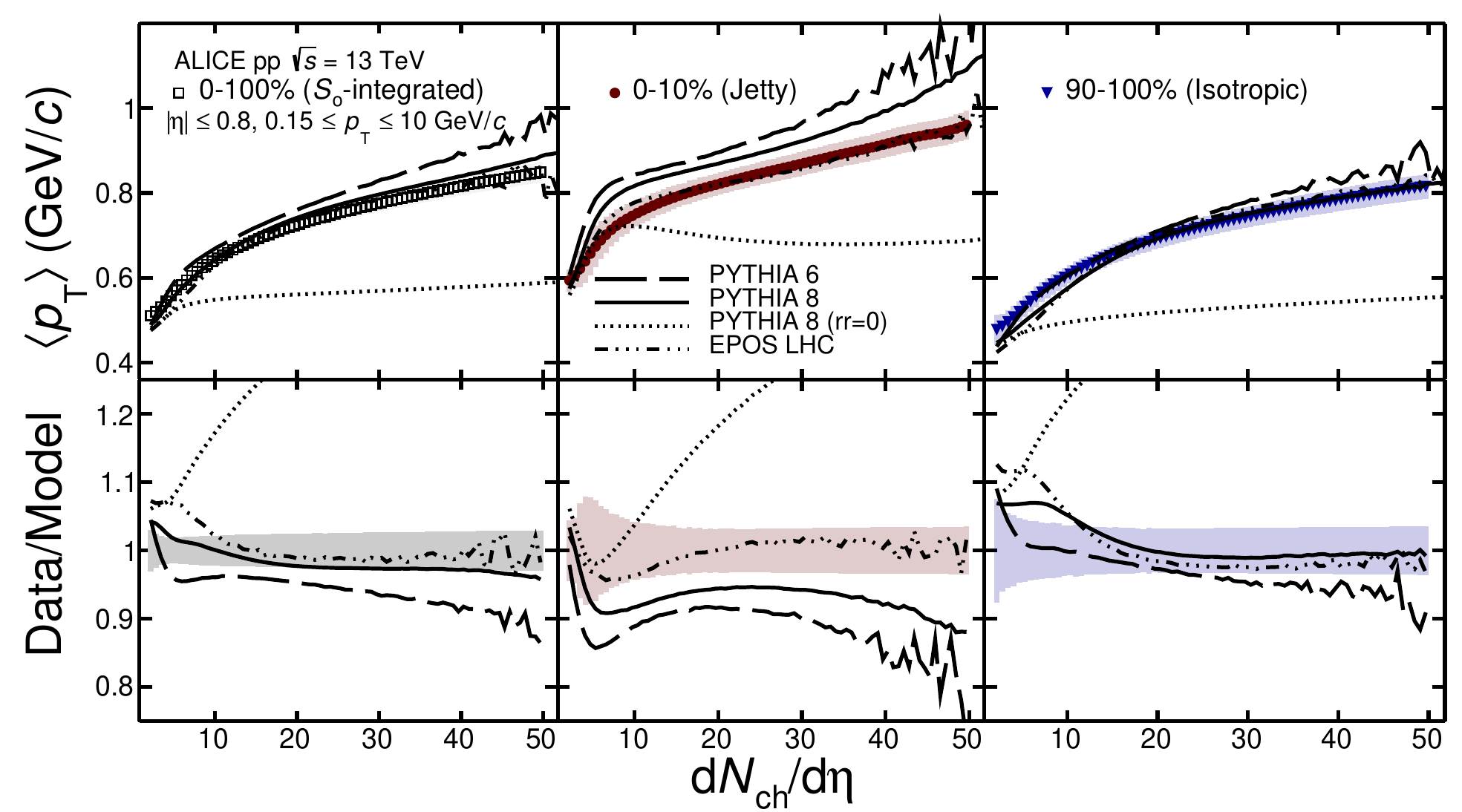}}
\caption{Average transverse momentum as a function of event multiplicity in pp collisions at $\sqrt{s}=13$\,TeV~\cite{Acharya:2019mzb}. Results for the spherocity-integrated case (0--100\%), the most jet-like (0--10\%) and isotropic (90--100\%) events are compared with Monte Carlo predictions. Predictions of PYTHIA~8 with and without (null reconnection range, $\rm{rr}=0$) color reconnection, as well as PYTHIA~6 and EPOS~LHC are displayed. Statistical uncertainties (error bars) are negligible compared to  systematic uncertainties (shaded area around the data points). Data to model ratios are shown in the bottom panel. The color band around unity represents the systematic uncertainty.}
\label{fig:5}
\end{figure}

The ALICE Collaboration has performed the strangeness analysis using pp data at $\sqrt{s}=13$\,TeV~\cite{Acharya:2019kyh}, and compared  data with MC event generators predictions.  PYTHIA 8.210~\cite{Sjostrand:2014zea}  (tune Monash~\cite{Skands:2014pea}) gives a reasonable description of the ${\rm K^{0}_S}$ results,  however it shows a less pronounced increase of the strange baryon yields with increasing \nch than what is observed in the data. Another generator is DIPSY~\cite{Flensburg:2011kk} which incorporates the so-called ``rope hadronization''. It does a better job in the description of data than PYTHIA, however, the discrepancy for the $\Omega$ distributions remains large. The EPOS-LHC~\cite{Werner:2007bf} model yields similar conclusions. The results suggest that the origin of the strangeness enhancement in hadronic collisions and its relation to the QCD deconfinement phase transition are still open problems for models.

Given the strong correlation between multiplicity and particle production at high \pt observed in pp data~\cite{Acharya:2019mzb}, the identified \pt spectra at high \nch have a non-negligible  contribution from  hard physics. In order to disentangle the soft and hard contribution, transverse spherocity was explored by ALICE. Spherocity is defined for a unit vector $\mathbf{ \hat{\rm \mathbf{n}}_{\rm \mathbf{s}} }$ which minimizes (min) the ratio:

\begin{equation}
S_{\rm 0} \equiv  \frac{\pi^{2}}{4}  \underset{\bf \hat{n}_{\rm \bf{s}}}{\rm min}  \left( \frac{\sum_{i}|{\vec p}_{{\rm T},i} \times { \bf \hat{n}_{\rm \bf{s}} }|}{\sum_{i}p_{{\rm T},i}}  \right)^{2},
\end{equation}
where the sum runs over all primary charged particles. At least three particles are required within $|\eta|<0.8$ and $\pt>0.15$\,GeV/$c$.  It is worth mentioning some features of spherocity:
\begin{itemize}
\item Low values of spherocity ($S_{0}\rightarrow0$) correspond to jet-like event topologies.
\item High values of spherocity ($S_{0}\rightarrow1$) correspond to event topologies where transverse momentum vectors are ``isotropically'' distributed.
\end{itemize}

ALICE measured the average transverse momentum as a function of multiplicity for different spherocity classes~\cite{Acharya:2019mzb}. Figure~\ref{fig:5} shows the average \pt as a function of ${\rm d}\nch/{\rm d}\eta$ for pp collisions at $\sqrt{s}=13$\,TeV.  In accordance with measurements at lower energies [21], the \mpt increases with ${\rm d}\nch/{\rm d}\eta$. The minimum bias data (spherocity-integrated event class) are compared with analogous measurements for the most jet-like structure (0--10\%) and isotropic (90--100\%) event classes.   Studying observables as a function of spherocity reveals interesting features.  On one hand, for isotropic events the average \pt is systematically below the spherocity-integrated \mpt over the full multiplicity range; on the other hand, for jet-like events the \mpt is higher than that for spherocity-integrated events. Within uncertainties, PYTHIA~8 with color reconnection gives an adequate description of the spherocity-integrated event class. It is worth mentioning that color reconnection was originally introduced to explain the rise of \mpt with multiplicity~\cite{Sjostrand:1987su}, and it turned into a ``QCD generator'' of radial flow-like patterns~\cite{Ortiz:2013yxa}. However, PYTHIA~6 shows a steeper rise of \mpt with ${\rm d}N_{\rm ch}/{\rm d}\eta$ than that seen in data. The comparison of data with EPOS~LHC is also shown. Clearly, the quantitative agreement is as good as that achieved by PYTHIA~8.  For the jetty-like and isotropic events corresponding to 0--10\% and 90--100\% spherocity classes, respectively, Fig.~\ref{fig:5} also shows comparisons between data and Monte Carlo generators (PYTHIA~6, PYTHIA~8 and EPOS~LHC). In low-multiplicity events (${\rm d}N_{\rm ch}/{\rm d}\eta<10$), the deviations between data and PYTHIA~8 (without color reconnection) are smaller and larger respectively for the 0--10\% and 90--100\% spherocity classes than those seen for the 0--100\% spherocity class. The effect could be a consequence of the reduction of the color reconnection contribution in events containing jets surrounded by a small underlying event activity. For isotropic events the three models quantitatively describe the correlation. Even for PYTHIA~6, the size of the discrepancy which was pointed out for the spherocity-integrated event class is reduced. On the contrary, for jet-like events both PYTHIA~6 and 8 exhibit a larger disagreement with the data.

Last but not least, it is worth mentioning that another approach is under investigation. It is inspired on results from heavy-ion collisions, where particle production is studied in both the jet-like and bulk regions, separately. Namely, the proton-to-pion ratio as a function of \pt in the bulk region exhibits a bump structure at intermediate \pt~\cite{Veldhoen:2012ge}. No bump is observed in the jet-like region. Therefore, one can argue that the bump observed in central heavy-ion collisions is a medium-related effect. Similar results for the Lambda-to-kaon ratio were reported for p--Pb collisions~\cite{Richert:2016bwd}. For pp collisions, an analogous separation implies the extension of the traditional underlying-event analysis~\cite{Field:2002vt,Martin:2016igp,Ortiz:2017jaz,Ortiz:2018vgc}. 

\section{Multi-particle correlations}
The key question which can be addressed using multi-particle correlations is whether the ridge phenomenon reflects initial momentum correlations from gluon saturation, or a final-state hydrodynamic response to the initial transverse collision geometry. The multi-particle cumulants probe the event-by-event  fluctuation of a single flow harmonic $v_{n}$, as well as the correlated fluctuations between the flow harmonics $v_{n}$ and $v_{m}$. As discussed above, the main challenge in small systems is how to disentangle the long-range ridge from ``non-flow'' correlations involving only few particles such as resonance decays or jets. For multi-particle cumulants, the non-flow contributions can be controlled by requiring correlations between particles from different subevents separated in $\eta$.

For example, for the recent results reported by ATLAS, the standard cumulant method consists on calculating the $k$-particle azimuthal correlations~\cite{Borghini:2001vi,Bilandzic:2010jr}, defined as:
\begin{equation}
\langle \{ 2 \}_{n} \rangle=\langle e^{in(\phi_{1}-\phi_{2})} \rangle,\,\langle \{ 3 \}_{n} \rangle=\langle e^{in(\phi_{1}+\phi_{2}-2\phi_{3})} \rangle,\,\langle \{ 4 \}_{n,m} \rangle=\langle e^{in(\phi_{1}-\phi_{2})+im(\phi_{3}-\phi_{4})} \rangle
\end{equation}
Tracks within the entire acceptance of $|\eta| < \eta_{\rm max} = 2.5$ are selected. The multi-particle asymmetric and symmetric cumulants are obtained as:
\begin{equation}
{\rm ac}_{n} \{ 3 \}=\langle\langle \{ 3 \}_{n} \rangle\rangle,\,{\rm sc}_{n,m}\{ 4 \}=\langle\langle \{ 4 \}_{n,m} \rangle\rangle - \langle\langle \{2\}_{n} \rangle\rangle  \langle\langle \{2\}_{m} \rangle\rangle,
\end{equation}
where $\langle\langle \rangle\rangle$ represents a weighted average of $\rm \langle \{ k \} \rangle$ over an event sample with similar multiplicity. In the absence of non-flow correlations, ${\rm sc}_{n,m}\{ 4 \}$ and ${\rm ac}_{n}\{ 3 \}$ measure the correlation between $v_{n}$ and $v_{m}$ or between $v_{n}$ and $v_{2n}$. Having said this, a technique to suppress further the non-flow correlations that typically involve a few particles within a localised region in $\eta$, has to be applied. Tracks are divided into several subevents, each covering a unique $\eta$ interval. The multi-particle correlations are then constructed by only correlating tracks between different subevents. In the two-subevent cumulant method, the tracks are divided into two subevents, labelled by $a$ and $b$, according to $-\eta_{\rm max} < \eta_{a} < 0$ and $0 < \eta_{b} < \eta_{\rm max}$. In the three-subevent cumulant method, tracks in each event are divided into three subevents $a$, $b$ and $c$, each covering one third of the $\eta$ range, $-\eta_{\rm max} < \eta_{a} < -\eta_{\rm max}/3$, $|\eta_{b}|\leq\eta_{\rm max}/3$  and $\eta_{\rm max}/3 < \eta_{c} < \eta_{\rm max}$. It has been shown that methods with three or more subevents are sufficient to reject now-flow correlation from jets.

\begin{figure}[htb]
\centerline{
\includegraphics[width=1.0\columnwidth]{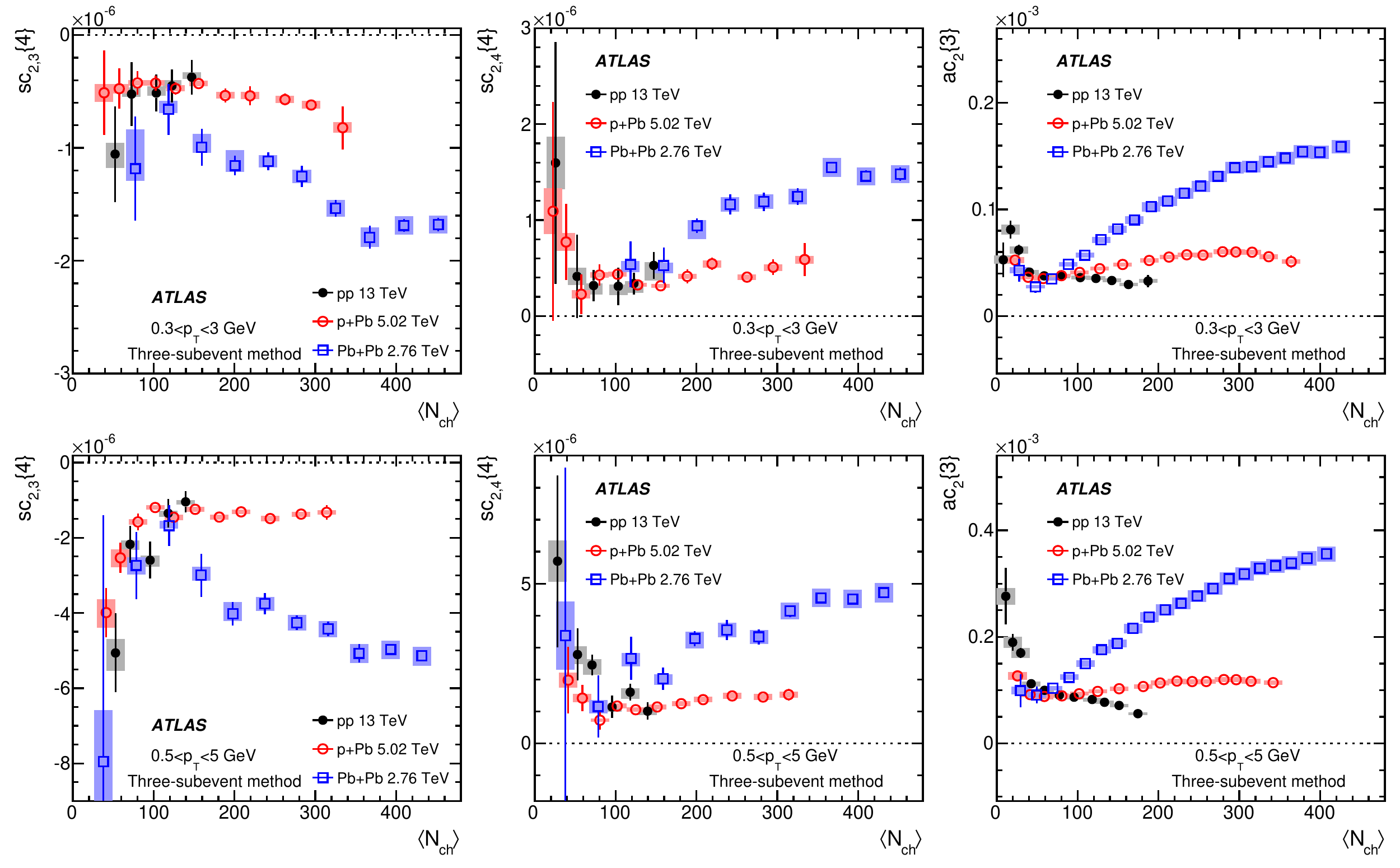}}
\caption{Multiplicity dependence of (left) ${\rm sc}_{2,3}\{4\}$, (middle) ${\rm sc}_{2,4}\{4\}$ and (right) ${\rm ac}_{2}\{3\}$ in (top) $0.3<\pt<3$\,GeV/$c$ and (bottom) $0.5<\pt<5$\,GeV/$c$ obtained for pp collisions (solid circles), p--Pb collisions (open circles) and low-multiplicity Pb--Pb collisions (open squares)~\cite{Aaboud:2018syf}. The error bars and shaded boxes represent the statistical and systematic uncertainties, respectively.}
\label{fig:2}
\end{figure}

Figure~\ref{fig:2} shows a comparison among results for pp, p--Pb and Pb--Pb collisions at $\sqrt{s}_{\rm NN}=13$, 5.02 and 2.76\,TeV, respectively~\cite{Aaboud:2018syf}.  Symmetric (${\rm sc}_{2,3}\{4\}$, ${\rm sc}_{2,4}\{4\}$) and asymmetric (${\rm ac}_{2}\{3\}$) cumulants,  for $0.3 < \pt < 3$\,GeV/$c$, are shown in the  top panel. These results show a negative correlation between $v_{2}$ and $v_{3}$ and a positive correlation between $v_2$ and $v_4$.  It is worth noticing that for the first time these patterns in small systems are observed, once  non-flow effects are adequately suppressed. In the multiplicity range covered by the pp collisions, $\langle N_{\rm ch} \rangle < 150$, the results for symmetric cumulants ${\rm sc}_{2,3} \{4\}$ and ${\rm sc}_{2,4} \{4\}$ are similar among the three systems. In the range $\langle N_{\rm ch} \rangle > 150$, $|{\rm sc}_{2,3} \{4\}|$ and ${\rm sc}_{2,4} \{4\}$ are larger in Pb--Pb than in p--Pb collisions. The results for ${\rm ac}_{2}\{3\}$ are similar among the three systems at $\langle N_{\rm ch} \rangle < 100$, but they deviate from each other at higher $\langle N_{\rm ch} \rangle$. The pp data are approximately constant or decrease slightly with $\langle N_{\rm ch} \rangle$, while the p--Pb and Pb--Pb data show significant increases as a function of $\langle N_{\rm ch} \rangle$. The bottom row shows the results for the higher \pt range of $0.5 < \pt < 5$ GeV/$c$, where similar trends are observed. Finally, for other collision systems such as Xe--Xe and Pb--Pb collisions at $\sqrt{s}_{\rm NN}=5.44$ and 5.02\,TeV, respectively, a positive ${\rm sc}_{2,4}$  was observed~\cite{Acharya:2019vdf}. While for ${\rm sc}_{2,4}$, results from ALICE, ATLAS and CMS qualitatively agree, the ALICE ${\rm sc}_{2,3}$ measurement goes from negative (anticorrelation between $v_2$ and $v_3$) to positive values at $N_{\rm ch} \sim100$. Same tendency is found for pp and p--Pb collisions. This different behavior is under investigation.

It has been shown that values of ${\rm sc}_{2,4}\{4\}$ and ${\rm ac}_2\{3\}$, which are both measures of correlations between $v_2$ and $v_4$, show significant differences between the standard method and the subevent methods. These differences seem to persist for $\langle N_{\rm ch} \rangle>200$ in p--Pb collisions and for $\langle N_{\rm ch} \rangle>150$ in Pb--Pb collisions, which is not compatible with the predicted behaviour of non-flow correlations at large $\langle N_{\rm ch} \rangle$~\cite{Aaboud:2018syf}. The effect may  arise from longitudinal flow decorrelations. The ATLAS and CMS collaborations measured decorrelation effects~\cite{Aaboud:2017tql,Khachatryan:2015oea}, which are large for $v_4$ and strongly correlated with $v_2$, and therefore they are expected to reduce the ${\rm sc}_{2,4}\{4\}$ and ${\rm ac}_{2}\{3\}$ in the subevent method.

\begin{figure}[htb]
\centerline{
\includegraphics[width=0.9\columnwidth]{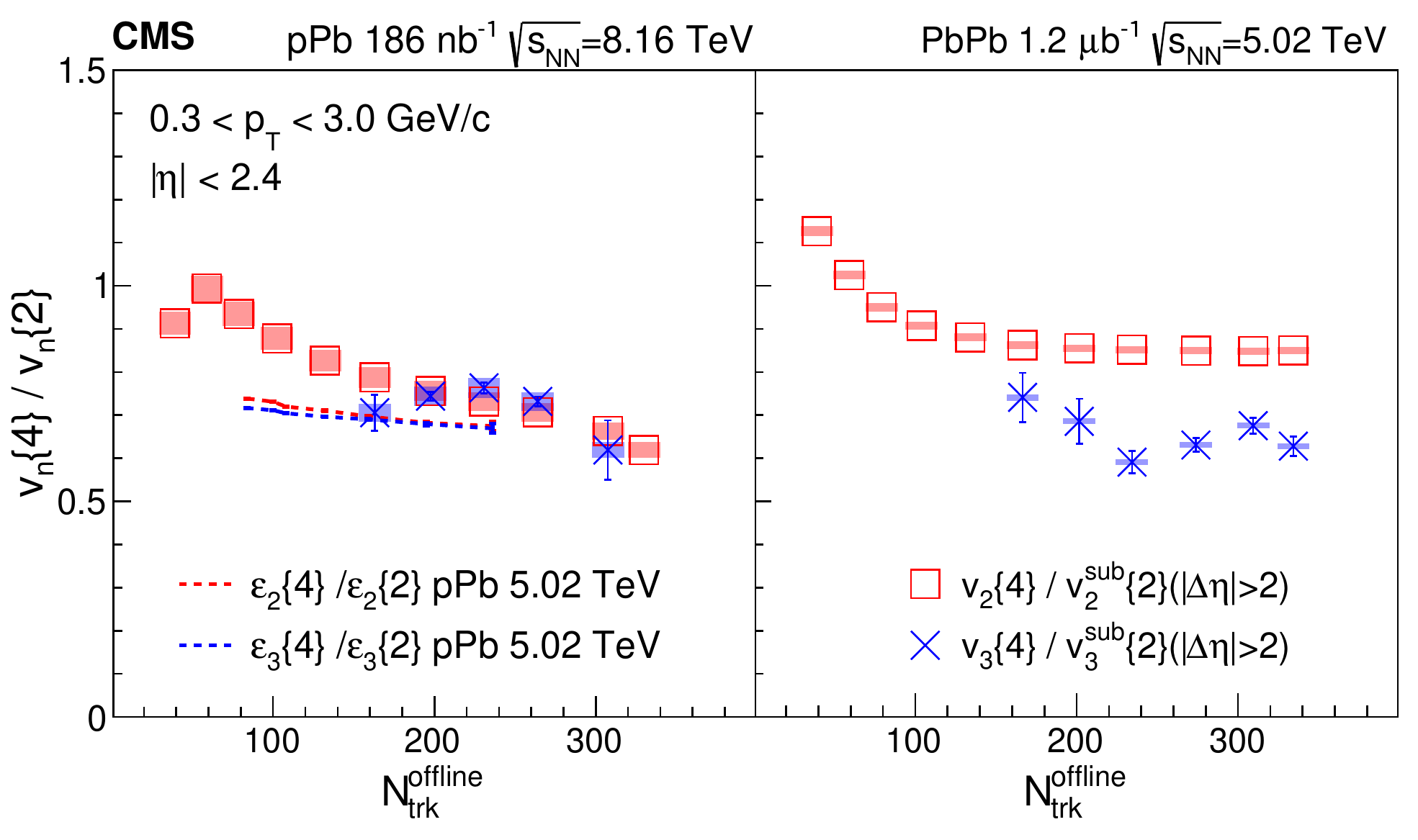}}
\caption{ The ratios of four- and two-particle harmonics ($v_2\{4\}/v2{2}$ and $v_3\{4\}/v_3\{2\}$) are shown as a function of event multiplicity for (left) p--Pb collisions at $\sqrt{s_{\rm NN}}=8.16$\,TeV and (right) Pb--Pb collisions at $\sqrt{s_{\rm NN}}=5.02$\,TeV~\cite{Sirunyan:2019pbr}. Error bars and shaded boxes denote statistical and systematic uncertainties, respectively.  The dashed curves show a hydrodynamics-motivated initial-state fluctuation calculation for p--Pb collisions at $\sqrt{s_{\rm NN}}=5.02$\,TeV.}
\label{fig:3}
\end{figure}

From the hydrodynamic picture, the $v_{n}\{m\}$ values are expected to be proportional to initial-state eccentricities $\epsilon_n\{m\}$~\footnote{Initial-state eccentricities are characterized by the cumulants of the event-by-event distributions of their Fourier harmonics coefficients.}, with $v_{n}\{m\}=k_{n} \epsilon_{n}\{m\}$, where $k_{n}$ reflects the medium properties and does not depend on the order of the cumulant.  Therefore, ratios of different cumulant $v_{n}$ values can directly probe properties of initial-state eccentricity. Using p--Pb data at $\sqrt{s_{\rm NN}}=8.16$\,TeV, the CMS collaboration has improved the precision on the measurement of the $v_{2}$ harmonics. Moreover, for the first time the $v_3$ harmonic was determined using multi-particle correlations~\cite{Sirunyan:2019pbr}. This allows the ratios $v_2\{4\}/v_2\{2\}$ and $v_3\{4\}/v_3\{2\}$ to be obtained in order to probe the properties of initial-state eccentricity. Figure~\ref{fig:3} shows the ratios $v_2\{4\}/v_2\{2\}$ and $v_3\{4\}/v_3\{2\}$ for both p--Pb and Pb--Pb collisions. For p--Pb collisions, the ratios for $v_2$ and $v_3$ are similar within uncertainties, which is consistent with having  both  the  second-  and  third-order  harmonics  arising  from  the  same  initial-state fluctuation mechanism. Comparing the p--Pb and Pb--Pb data, the $v_3$ ratios are comparable for both systems, while the $v_2$ ratios are higher in Pb--Pb than in p--Pb for higher multiplicity values, again reflecting the larger geometric contribution for the larger collision system. The figure also shows that a hydrodynamics-motivated initial-state fluctuation calculation of eccentricities for p--Pb collisions at $\sqrt{s_{\rm NN}}=5.02$\,TeV, qualitatively describes the data.

\section{Summary}

In this work, selected results on particle correlations and transverse momentum spectra in small systems have been presented. From the experimental side, new techniques were developed in order to improve the removal of trivial correlations from flow measurements. For example, using the sub-event method a non zero $v_2$ is observed for pp and p--Pb collisions in accordance with hydrodynamic predictions. Moreover, the anti-correlation between $v_2$ and $v_3$ (as expected from hydrodynamics) in small systems has been reported by ATLAS and CMS. However, ALICE data exhibit a different behavior for low multiplicity events, this difference needs to be fully understood. Regarding transverse momentum spectra as a function of multiplicity, radial flow-like behavior is observed in data, and now the role of autocorrelations is under investigation using an event selection based on transverse spherocity. New measurements of strange and multistrange particles as a function of multiplicity have been performed. The MC event generators can not fully describe the data.

We are still far from a complete understanding of the new phenomena discovered in small systems. More work needs to be done both from the theory and experimental sides, in order to find a unified description of the system size dependence of particle production.

\section{Acknowledgments}
The presenter has been supported by PAPIIT-UNAM  under the Grant No. IN102118.

\bibliographystyle{utphys}   
\bibliography{biblio}


\end{document}